\documentclass[12pt]{article}
\usepackage{amsmath}
\usepackage{amsfonts}
\usepackage{graphicx}
\usepackage{mathrsfs}

\begin{document}

%
\def\papertitlepage{\baselineskip 3.5ex \thispagestyle{empty}}
\def\preprinumber#1#2{\hfill \begin{minipage}{4.2cm}  #1
                 \par\noindent #2 \end{minipage}}
\renewcommand{\thefootnote}{\fnsymbol{footnote}}
\newcommand{\beq}{\begin{equation}}
\newcommand{\eeq}{\end{equation}}
\newcommand{\beqa}{\begin{eqnarray}}
\newcommand{\eeqa}{\end{eqnarray}}
\newcommand{\n}{\nonumber\\}
%
%
%
%
\papertitlepage
\setcounter{page}{0}
\preprinumber{KEK-TH-1509}{}
\baselineskip 0.8cm
\vspace*{2.0cm}
\begin{center}
{\large\bf Chiral Zeromodes
on Vortex-type \\Intersecting Heterotic Five-branes
}
\end{center}
\vskip 4ex
\baselineskip 1.0cm
\begin{center}
           {Masaya Yata} 
\\
\vskip 1em
    {\it 
Department of Particle and Nuclear Physics}\\
 \vskip -2ex {\it The Graduate University for Advanced Studies}\\
 \vskip -2ex {\it Tsukuba, Ibaraki 305-0801, Japan} \\

\end{center}
\vskip 5ex
%
\baselineskip=3.5ex
\begin{center} {\bf Abstract} \end{center}

\vskip 2ex

We solve the gaugino Dirac equation on a smeared intersecting five-brane solution in 
$E_8~\times~E_8$ heterotic string theory to search for localized chiral zeromodes 
on the intersection. 
The background is chosen to depend on the full two-dimensional overall 
transverse coordinates to the branes. Under some appropriate boundary conditions, 
we compute the complete spectrum of zeromodes to find that, among infinite towers of 
Fourier modes, there exist only three localized normalizable zeromodes, one of which 
has opposite chirality to the other two. This agrees with the result previously obtained in the 
domain-wall type solution, supporting the claim that there exists one net chiral zeromode localized 
on the heterotic five-brane system.

\vspace*{\fill}
\noindent
November 2011
\newpage
\renewcommand{\thefootnote}{\arabic{footnote}}
\setcounter{footnote}{0}
\setcounter{section}{0}
\baselineskip = 0.6cm
\pagestyle{plain}
%
Flux compactifications of string theory on non-K{\"a}hler manifolds 
\cite{Rohm:1985jv}-\cite{
Serone:2003sv} have attracted much interest for some recent years. 
In heterotic string theory, it has been known for a long time that the moduli are partially stabilized by the 
NS-NS three form flux $H_{MNP}$\cite{Rohm:1985jv}\cite{Strominger:1986uh}\cite{Dine:1985rz}. 
In general, turning on fluxes produces a potential \cite{Gukov:1999ya}, and one finds new vacua with some stabilized moduli at the potential minima \cite{Kachru:2003aw}-\cite{Gauntlett:2003cy}. 
More recently, 
understanding of heterotic moduli stabilization was improved by taking into account the intrinsic torsion 
of the geometry
\cite{Lopes Cardoso:2002hd}
-\cite{Serone:2003sv}.

Historically, heterotic string theory was extensively studied in 1980s as a candidate unified theory
including gravity. After the D-branes were found, however, the research focus 
has shifted from heterotic to type II theories, in particular,  while the brane-world scenario \cite{Randall:1999ee} \cite{Randall:1999vf}
was readily realized in type II theories using D-branes, there are no known 
such constructions 
in heterotic string theory. Despite this, an attempt was made in \cite{Kimura:2009tb} 
to realize warped compactification of heterotic string theory by using NS5-branes,
where  the
authors considered a domain-wall type smeared intersecting NS5-brane solution in $E_8~{\times}~E_8$ heterotic string theory
and explicitly solved the gaugino Dirac equation to find 
one net chiral fermionic zeromode. This result was in agreement with the naive 
counting argument of Nambu-Goldstone modes on this background \cite{Kimura:2009tb}.  

In this paper, 
we perform the gaugino-zeromode analysis on a similar smeared intersecting 
background, in which, unlike the one considered in \cite{Kimura:2009tb}, 
the field configurations depend on not only just one 
of the overall transverse coordinates (that is, the domain-wall type) but {\em full two}-dimensional 
overall transverse coordinates to the branes (the vortex type)
\footnote{They constitute the intersecting $p$-brane 
solution in the original form obtained in \cite{Argurio:1997gt,Ohta:1997gw}. Note that the relatively transverse dimensions are
still smeared. 
}.
%
Although 
the Dirac operator becomes nontrivial and much more 
complicated than the one considered in \cite{Kimura:2009tb},  
we will solve the zeromode equation under some boundary conditions, and 
compute the complete spectrum of zeromodes. 
In particular, we will find that, among infinite towers of Fourier modes,  
there exist only three localized normalizable zeromodes, one of which has opposite chirality 
to the other two. This agrees with the result obtained in \cite{Kimura:2009tb}, 
supporting the claim that there exists one net chiral zeromode localized 
on the intersection of the heterotic five-brane system.

\begin{figure}[t]
 \begin{minipage}{0.5\hsize}
  \begin{center}
   \includegraphics[scale=0.8]{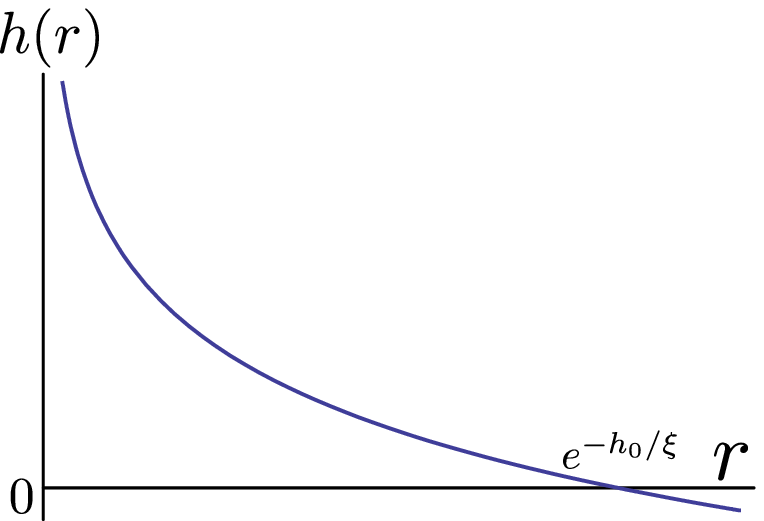}
  \end{center}
  \caption{$\xi<0$}
  \label{fig:one}
 \end{minipage}
 \begin{minipage}{0.5\hsize}
  \begin{center}
   \includegraphics[scale=0.8]{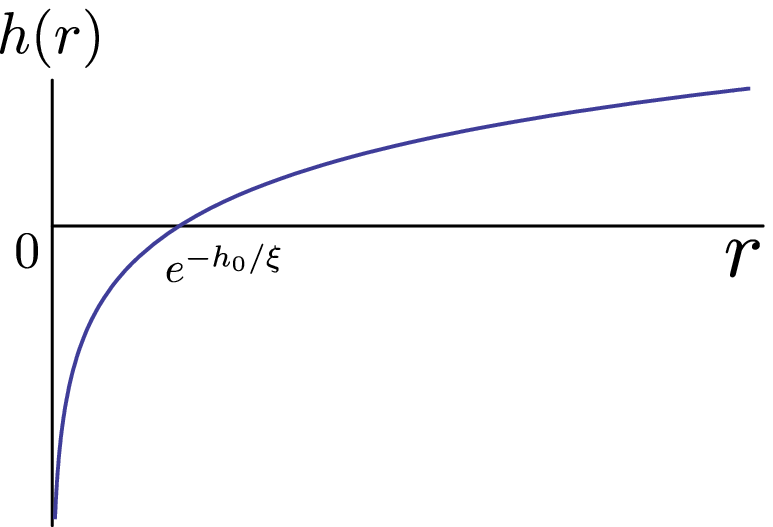}
  \end{center}
  \caption{$\xi>0$}
  \label{fig:two}
 \end{minipage}
\end{figure}

We begin with the neutral smeared intersecting five-brane solution \cite{Argurio:1997gt} \cite{Ohta:1997gw}:
\begin{eqnarray}
ds^2&=& \sum_{i,j=0,7,8,9}\eta_{ij}dx^idx^j+h(x^1,x^2)^2\sum_{\mu,\nu=1,2}\delta_{\mu\nu}dx^{\mu}dx^{\nu}+
                                          h(x^1,x^2)\sum_{\mu,\nu=3,4,5,6}\delta_{\mu\nu}dx^{\mu}dx^{\nu},\n
h(x^1,x^2)^2&=&e^{2\phi}, \label{Smeared Solution}
\end{eqnarray}
where
\begin{eqnarray}
h(x^1,x^2)=h_0+\xi\log{r}~,~~r=\sqrt{(x^1)^2+(x^2)^2}, \label{potential h}
\end{eqnarray}
$h_0$ and $\xi$ are real constants.
The profiles of the harmonic function $h(x^1,x^2)$ are shown in Figure \ref{fig:one}
for $\xi<0$, and in Figure \ref{fig:two} for $\xi>0$.
Since $h(x^1,x^2)$ is equal to the string coupling, we only consider the region where 
$h(x^1,x^2)$ is positive, and impose the boundary condition that 
all the fields become $0$ where $h=0$.
The 3-form flux $H_{\mu\nu\rho}$ is 
\begin{eqnarray}
H_{\mu\nu\rho}=
\begin{cases}
{1\over2}{\partial h(x^1,x^2)\over \partial x^1} & \mbox{if $(\mu,\nu,\rho)=(2,3,4),(2,5,6)$ and even permutations},\\
-{1\over2}{\partial h(x^1,x^2)\over \partial x^1}  & \mbox{if $(\mu,\nu,\rho)=(2,4,3),(2,6,5)$ and even permutations},\\
-{1\over2}{\partial h(x^1,x^2)\over \partial x^2}  & \mbox{if $(\mu,\nu,\rho)=(1,3,4),(1,5,6)$ and even permutations},\\
{1\over2}{\partial h(x^1,x^2)\over \partial x^2}  & \mbox{if $(\mu,\nu,\rho)=(1,4,3),(1,6,5)$ and even permutations},\\
0 & \mbox{otherwise}.
\end{cases} \label{H flux}
\end{eqnarray}
(\ref{Smeared Solution}) and (\ref{H flux}) are 
a solution to the equations of motion of the type II NS-NS sector Lagrangian.
The metric (\ref{Smeared Solution}) 
represents two NS5-branes extended in the dimensions shown in
Table \ref{5brane}.
We emphasize here that our solution is 
different from the one adopted in \cite{Kimura:2009tb} in that the harmonic function 
(\ref{potential h}) has a dependence on both the $x^1$ and $x^2$ coordinates. 
\newcommand{\bhline}[1]{\noalign{\hrule height #1}}
\newcommand{\bvline}[1]{\vrule width #1}
\begin{table}[b]
\begin{center}
\begin{tabular}{@{\bvline{1pt}}c|c|c|c|c|c|c|c|c|c|c@{\bvline{1pt}}}
 \bhline{1pt}
~~  & 0 & 1 & 2 & 3 & 4 & 5 & 6 & 7 & 8 & 9~ \\ \hline
~5-brane1 & ${\times}$ & ~ & ~ & ~ & ~ & ${\times}$ & ${\times}$ & ${\times}$ & ${\times}$ & ${\times}$~ \\ \hline
~5-brane2 & ${\times}$ & ~ & ~ & ${\times}$ & ${\times}$ & ~ & ~ & ${\times}$ & ${\times}$ & ${\times}$~ \\
\bhline{1pt}
\end{tabular}
\caption{Dimensions in which the 5-branes extend}\label{5brane}
\end{center}
\end{table}

In (\ref{potential h}), the parameter $\xi$ is related to the NS5-brane tension. 
The brane action is calculated by using 
the equations of  motion as follows (in the Einstein frame):
\begin{eqnarray}
S_{\mbox{\tiny 5-Brane1}}=
-{T\over2\kappa^2}\int d^6x\sqrt{-\det{G_{mn}}}e^{-\phi/2}\delta(x^1,x^2), \nonumber
\end{eqnarray}
where $G_{mn}$ is the six-dimensional metric $(m,n=0,5,6,7,8,9)$,
 and $T$ is the brane tension
\begin{eqnarray}
T=-2\pi\xi. \nonumber
\end{eqnarray}
Therefore, if $\xi < 0$, the 5-brane tension is positive. 
On the other hand, if $\xi >0$, the tension becomes negative, implying that 
this is an orientifold-like object.
In \cite{Imazato:2011mj}, it was argued that such an object  in heterotic string theory 
might be understood as a mirror to the Atiyah-Hitchin manifold.
In this paper, we only study positive tension branes.


Next we convert this type II NS5-brane background into an $E_8~{\times}~E_8$ heterotic 
background by the standard embedding \cite{Callan:1991dj}.
We generalize the spin connection $\omega_{\mu}$ by adding the 3-form flux $H_{\mu\nu\rho}$
\begin{eqnarray}
\Omega_{\pm\mu}^{~\alpha\beta}=\omega_{\mu}^{~\alpha\beta}\pm H_{\mu}^{~\alpha\beta},
\end{eqnarray}
and we identify $\Omega_{+\mu}$ to the gauge connection $A_{\mu}$. 
We present the gauge connections using Gell-Mann matrices $\lambda_i$ $(i=1,{\cdots},8)$ and $2{\times}2$ matrices 
$
\mbox{\boldmath $1$}\equiv
\left(
\begin{array}{cc}
1 & 0\\
0 & 1
\end{array}
\right)
$, 
$
\mbox{\boldmath $s$}\equiv i\sigma_2=
\left(
\begin{array}{cc}
0 & 1 \\
-1 & 0
\end{array}
\right)
$:
\begin{eqnarray}
A_1^{~\alpha\beta}&=&
-{1\over h(x^1,x^2)}{\partial h(x^1,x^2)\over \partial x^2}
\{
-{3\over4}(3\lambda_3+\sqrt{3}\lambda_8)\}~{\otimes}~\mbox{\boldmath $s$}, \label{gauge A}\\
A_2^{~\alpha\beta}&=&
{1\over h(x^1,x^2)}{\partial h(x^1,x^2)\over \partial x^1} 
\{
-{3\over4}(3\lambda_3+\sqrt{3}\lambda_8)\}~{\otimes}~\mbox{\boldmath $s$}, \n
A_3^{~\alpha\beta}&=&
{1\over2 h(x^1,x^2)^{3/2}}{\partial h(x^1,x^2)\over \partial x^1}
(-i\lambda_2)~{\otimes}~\mbox{\boldmath $1$ } -
{1\over2 h(x^1,x^2)^{3/2}}{\partial h(x^1,x^2)\over \partial x^2}
(-\lambda_1)~{\otimes}~\mbox{\boldmath $s$}, \n
A_4^{~\alpha\beta}&=&
{1\over2 h(x^1,x^2)^{3/2}}{\partial h(x^1,x^2)\over \partial x^1}
(-\lambda_1)~{\otimes}~\mbox{\boldmath $s$ } +
{1\over2 h(x^1,x^2)^{3/2}}{\partial h(x^1,x^2)\over \partial x^2}
(-i\lambda_2)~{\otimes}~\mbox{\boldmath $1$}, \n
A_5^{~\alpha\beta}&=&
{1\over2 h(x^1,x^2)^{3/2}}{\partial h(x^1,x^2)\over \partial x^1}
(-i\lambda_5)~{\otimes}~\mbox{\boldmath $1$ } +
{1\over2 h(x^1,x^2)^{3/2}}{\partial h(x^1,x^2)\over \partial x^2}
(+\lambda_4)~{\otimes}~\mbox{\boldmath $s$}, \n
A_6^{~\alpha\beta}&=&
-{1\over2 h(x^1,x^2)^{3/2}}{\partial h(x^1,x^2)\over \partial x^1}
(+\lambda_4)~{\otimes}~\mbox{\boldmath $s$ } -
{1\over2 h(x^1,x^2)^{3/2}}{\partial h(x^1,x^2)\over \partial x^2}
(-i\lambda_5)~{\otimes}~\mbox{\boldmath $1$}. \nonumber
\end{eqnarray}
$\Omega_{+\mu}$, which is generically an $SO(6)$ spin connection, can be written 
in this form and set equal to the gauge connection $A_{\mu}$, thanks to 
the $SU(3)$ structure of this background. 
The eigenvalues of $\mbox{\boldmath $s$}$, $\pm i$, distinguish which $SU(3)$ representation 
the gaugino is in, $\mbox{\boldmath $3$}$ or $\bar{\mbox{\boldmath $3$}}$. 
We adopt $\mbox{\boldmath $s$}=+i$ from now on. 
These backgrounds (\ref{H flux}) and (\ref{gauge A}) preserve 1/4 of supersymmetries since
the generalized spin connections $\Omega_{\pm\mu}$ are in $SU(3)$. 
In order to satisfy the Bianchi identity $dH=0$, we embed the gauge group $SU(3)$ 
to $E_8~{\times}~E_8$ and get the unbroken gauge symmetry $E_6(\times E_8)$. 
The adjoint representation of $E_8$ is decomposed by embedding $SU(3)$ as follows: 
\begin{eqnarray}
\mbox{\boldmath $248$}=(\mbox{\boldmath $78$},\mbox{\boldmath $1$})~{\oplus}~
                       (\mbox{\boldmath $27$},\mbox{\boldmath $3$})~{\oplus}~
                       (\bar{\mbox{\boldmath $27$}},\bar{\mbox{\boldmath $3$}})~{\oplus}~
                       (\mbox{\boldmath $1$},\mbox{\boldmath $8$}).
\end{eqnarray}
Since the gauge field has a vev in $SU(3)$, 
the fields which are transformed as $(\mbox{\boldmath $27$},\mbox{\boldmath $3$})~{\oplus}~(\bar{\mbox{\boldmath $27$}},\bar{\mbox{\boldmath $3$}})$ (as well as $ (\mbox{\boldmath $1$},\mbox{\boldmath $8$})$) become Nambu-Goldstone bosons. 
We know that a $D=4$, ${\cal N}=1$ chiral supermultiplet have only two bosonic degrees of freedom. 
The Nambu-Goldstone bosons, which belongs to $\mbox{\boldmath $27$}$ and $\bar{\mbox{\boldmath $27$}}$, must be combined with their superpartners 
to a single ${\cal N}=1$ chiral supermultiplet. 
This means that the moduli are a triplet (of the broken $SU(3)$) of chiral supermultiplets that transform as $\mbox{\boldmath $27$}$ (or $\bar{\mbox{\boldmath $27$}}$)  of $E_6$. 
Therefore, from this bosonic moduli counting argument  
one might conclude 
that there would be three chiral zeromodes. 
In fact, however, there is left only one net generation of fermions on the intersecting NS5-brane 
since the chiralities of the localized solutions are different, as we show below. 
The results agree with \cite{Kimura:2009tb}.

To know the number of generations that localize on the branes, 
we need to compute the Dirac indices. 
The ten-dimensional heterotic gaugino equation of motion is
\begin{eqnarray}
\Gamma^M D_M(\omega-{1\over3}H,A)\chi-\Gamma^M\chi\partial_M\phi
+{1\over8}\Gamma^M\gamma^{AB}(F_{AB}+\hat{F}_{AB})(\psi_M+{2\over3}\Gamma_M\lambda)=0,
\end{eqnarray}
where
\begin{eqnarray}
D_M(\omega-{1\over3}H,A)\chi 
\equiv\Bigl( \partial_M + {1\over4}(\omega_M^{~AB}-{1\over3}H_M^{~AB})\gamma_{AB}+\mbox{ad}A_M \Bigr)\chi
\end{eqnarray}
and $\mbox{ad}A_M{\cdot}\chi\equiv[A_M,\chi]$.
We set $\psi_M=0$, $\lambda=0$ and $\tilde{\chi}\equiv e^{-\phi}\chi$, the equation of motion becomes
\begin{eqnarray}
\Gamma^MD_M(\omega-{1\over3}H,A)\tilde{\chi}=0. \label{10D Dirac}
\end{eqnarray}
The $SO(9,1)$ gamma matrices $\Gamma^M$ are
\begin{eqnarray}
\Gamma^a=\gamma^a_{\mbox{\tiny 4D}}{\otimes}\mbox{\boldmath $1$}_8,~~(a=0,7,8,9) ~~~
\Gamma^{\alpha}=\gamma^{\sharp}_{\mbox{\tiny 4D}}{\otimes}\gamma^{\alpha}~~(\alpha=1,{\dotsc},6) \nonumber
\end{eqnarray}
where $\gamma^a_{\mbox{\tiny 4D}}$, $\gamma^{\sharp}_{\mbox{\tiny 4D}}$ are the ordinary $SO(3,1)$ gamma 
matrices and chiral operator, respectively.
$\gamma^{\alpha}$ are the $SO(6)$ gamma matrix and we fix them as
\beqa
\gamma^1=\sigma_1{\otimes}\mbox{\boldmath$1$}{\otimes}\mbox{\boldmath$1$},~~\gamma^2=\sigma_2{\otimes}\mbox{\boldmath$1$}{\otimes}\mbox{\boldmath$1$},~~\gamma^3=\sigma_3{\otimes}\sigma_1{\otimes}\mbox{\boldmath$1$}, \nonumber\\
\gamma^4=\sigma_3{\otimes}\sigma_2{\otimes}\mbox{\boldmath$1$},~~\gamma^5=\sigma_3{\otimes}\sigma_3{\otimes}\sigma_1,~~\gamma^6=\sigma_3{\otimes}\sigma_3{\otimes}\sigma_2 .\nonumber
\eeqa
The $SO(6)$ chiral operator is defined by $\gamma^{\sharp}=-i\gamma^1\gamma^2\gamma^3\gamma^4\gamma^5\gamma^6$.
More explicitly, it is represented in the matrix form:
\beqa
\gamma^{\sharp}
=
\left(
\begin{array}{cccccccc}
-\mbox{\boldmath $1$}_3 & 0 & 0 & 0 & 0 & 0 & 0 & 0 \\
 0 & \mbox{\boldmath $1$}_3 & 0 & 0 & 0 & 0 & 0 & 0 \\
 0 & 0 & \mbox{\boldmath $1$}_3 & 0 & 0 & 0 & 0 & 0 \\
 0 & 0 & 0 &-\mbox{\boldmath $1$}_3 & 0 & 0 & 0 & 0 \\
 0 & 0 & 0 & 0 & \mbox{\boldmath $1$}_3 & 0 & 0 & 0 \\
 0 & 0 & 0 & 0 & 0 &-\mbox{\boldmath $1$}_3 & 0 & 0 \\
 0 & 0 & 0 & 0 & 0 & 0 &-\mbox{\boldmath $1$}_3 & 0 \\
 0 & 0 & 0 & 0 & 0 & 0 & 0 & \mbox{\boldmath $1$}_3
\end{array}
\right) \label{chiral op}
\eeqa
where $\mbox{\boldmath $1$}_3=
\left(
\begin{array}{ccc}
1&0&0\\
0&1&0\\
0&0&1
\end{array}
\right)
$.
We can decompose an $SO(9,1)$ Majorana-Weyl spinor into $SO(3,1)$ and $SO(6)$ spinors as 
$\mbox{\boldmath $16$} =(\mbox{\boldmath $2$}_+,\mbox{\boldmath $4$}_+)~{\oplus}~(\mbox{\boldmath $2$}_-,\mbox{\boldmath $4$}_-)$, 
where the subscripts $\pm$ are the $SO(3,1)$ and $SO(6)$ chiralities. 
Since the $SO(9,1)$ spinor is Majorana, 
the $(\mbox{\boldmath $2$}_+,\mbox{\boldmath $4$}_+)$ and $(\mbox{\boldmath $2$}_-,\mbox{\boldmath $4$}_-)$ components are associated with charge conjugation.

Equation (\ref{10D Dirac}) can be divided into $i=0,7,8,9$ and $\mu =1,{\cdots},6$ directions \cite{Kimura:2006af}
\begin{eqnarray}
\Gamma^i\partial_i \tilde{\chi}+
\Gamma^{\mu}D_{\mu}(\omega-{1\over3}H,A)\tilde{\chi}=0.
\end{eqnarray}
If $\tilde{\chi}=\tilde{\chi}_{\mbox{\tiny 4D}}~{\otimes}~\tilde{\chi}_{\mbox{\tiny 6D}}$, 
the second term looks like the mass term of the four dimensional Dirac equation.
Being a singular noncompact geometry, no index theorem is available on our background. Therefore,  in order to count the number of localized fermionic zeromodes 
on the intersection of the five-branes 
we will solve the Dirac equation directly
\begin{eqnarray}
\gamma^{\mu}D_{\mu}(\omega-{1\over3}H,A)\tilde{\chi}_{\mbox{\tiny 6D}}=0 \label{6D Dirac}
\end{eqnarray}
and find localized solutions which may have either positive or negative chirality.

The gauge connection $A_{\mu}$ has nonzero value in the $SU(3)$ subalgebra, 
and therefore the $\tilde{\chi}_{\mbox{\tiny 6D}}$ is transformed as a triplet of $SU(3)$. 
We know that the chiralities of $\tilde{\chi}_{\mbox{\tiny 6D}}$ related to each other by charge conjugation. 
If we change the representation of the $SU(3)$, $\mbox{\boldmath $3$}$ or $\bar{\mbox{\boldmath $3$}}$, 
the chirality of $\tilde{\chi}_{\mbox{\tiny 6D}}$ also changes oppositely.  
Since $\tilde{\chi}_{\mbox{\tiny 6D}}$ depend only  $x^1$ and $x^2$, (\ref{6D Dirac}) becomes
\begin{eqnarray}
\left(
\begin{array}{cc}
     ~         & \partial_{\bar{z}}~\mbox{\boldmath $1$}_{12} \\
\partial_{z}~\mbox{\boldmath $1$}_{12}   &         ~ 
\end{array}
\right)
\tilde{\chi}_{\mbox{\tiny 6D}}+
{\cal{M}}
\tilde{\chi}_{\mbox{\tiny 6D}}=0, \label{new 6D Dirac}
\end{eqnarray}
where 
\begin{eqnarray}
\partial_{z}&=&{\partial \over \partial z}\equiv
{\partial \over \partial x^1} + i {\partial \over \partial x^2},~~
\partial_{\bar{z}}={\partial \over \partial \bar{z}}\equiv
{\partial \over \partial x^1} - i {\partial \over \partial x^2}, \nonumber\\
{\cal M} &=& 
\gamma^{\mu}\{{1\over4}(\omega_{\mu}^{~AB}-{1\over3}H_{\mu}^{~AB})\Gamma_{AB}+\mbox{ad}A_{\mu}\}, \nonumber
\end{eqnarray}
and ${\mbox{\boldmath $1$}}_{12}$ is the $12{\times}12$ unit matrix.
$\tilde{\chi}_{\mbox{\tiny 6D}}$ is a fermion that has 24 components, and
we solve the Dirac equation (\ref{new 6D Dirac}) for each component.
In order to find solutions, we expand these components by Fourier modes as 
\begin{eqnarray}
\tilde{\chi}_{\mbox{\tiny 6D}}^{\pm N}=\sum_{m=-\infty}^{\infty}e^{im\theta}\tilde{\chi}_{\mbox{\tiny 6D}~m}^{\pm N}(r),
\end{eqnarray}
where the signs $\pm$ denote the chiralities, and $N$ $(=1,{\cdots},12)$ label the 
components of the gaugino $\tilde{\chi}_{\mbox{\tiny 6D}}$.
Thus we obtain 24 Dirac equations from (\ref{new 6D Dirac}) form each component, 
some of which are complicated 
differential equations.
However, we can find a linear transformation matrix $T$ :
\begin{eqnarray}
T=
\left(
\begin{array}{cc}
t_1 & t_2 \\
t_2 & t_1
\end{array}
\right), \nonumber
\end{eqnarray}
\begin{eqnarray}
t_1=
\left(
\begin{array}{cccccccccccc}
  0  &  0  &  0  &  0  &  0  &-1/2 &  0  &-1/2 &  0  &  0  &  0  &  0  \\
  0  &  0  &  0  &  0  &  0  &  0  &  0  &  0  &  0  &  0  &  0  &  0  \\
  0  &  0  &  0  &  0  &  0  &  0  &  0  &  0  &  0  &  0  &  0  &  0  \\
  0  &  0  &  0  & 1/2 &  0  &  0  &  0  &  0  &  0  &  0  &  0  &  0  \\
  0  &  0  &  0  &  0  &  1  &  0  &  0  &  0  &  0  &  0  &  0  &  0  \\
  0  &  0  &  0  &  0  &  0  & 1/2 &  0  &-1/2 &  0  &  0  &  0  &  0  \\
  0  &  0  &  0  &  0  &  0  &  0  & 1/2 &  0  &  0  &  0  &  0  &  0  \\
  0  &  0  &  0  &  0  &  0  & 1/4 &  0  & 1/4 &  0  &  0  &  0  &  0  \\
  0  &  0  &  0  &  0  &  0  &  0  &  0  &  0  &  1  &  0  &  0  &  0  \\
  0  &  0  &  0  &  0  &  0  &  0  &  0  &  0  &  0  &  0  &  0  &  0  \\
  0  &  0  &  0  & 1/2 &  0  &  0  &  0  &  0  &  0  &  0  &  0  &  0  \\
  0  &  0  &  0  &  0  &  0  &  0  & 1/2 &  0  &  0  &  0  &  0  &  0 
\end{array}
\right), \nonumber
\end{eqnarray}
\begin{eqnarray}
t_2=
\left(
\begin{array}{cccccccccccc}
 1/2 &  0  &  0  &  0  &  0  &  0  &  0  &  0  &  0  &  0  &  0  &  0  \\
  0  &  1  &  0  &  0  &  0  &  0  &  0  &  0  &  0  &  0  &  0  &  0  \\
  0  &  0  &  1  &  0  &  0  &  0  &  0  &  0  &  0  &  0  &  0  &  0  \\
  0  &  0  &  0  &  0  &  0  &  0  &  0  &  0  &  0  &  0  &-1/2 &  0  \\
  0  &  0  &  0  &  0  &  0  &  0  &  0  &  0  &  0  &  0  &  0  &  0  \\
  0  &  0  &  0  &  0  &  0  &  0  &  0  &  0  &  0  &  0  &  0  &  0  \\
  0  &  0  &  0  &  0  &  0  &  0  &  0  &  0  &  0  &  0  &  0  &-1/2 \\
 1/2 &  0  &  0  &  0  &  0  &  0  &  0  &  0  &  0  &  0  &  0  &  0  \\
  0  &  0  &  0  &  0  &  0  &  0  &  0  &  0  &  0  &  0  &  0  &  0  \\
  0  &  0  &  0  &  0  &  0  &  0  &  0  &  0  &  0  &  1  &  0  &  0  \\
  0  &  0  &  0  &  0  &  0  &  0  &  0  &  0  &  0  &  0  & 1/2 &  0  \\
  0  &  0  &  0  &  0  &  0  &  0  &  0  &  0  &  0  &  0  &  0  & 1/2
\end{array}
\right).\nonumber
\end{eqnarray}
%
and define a new gaugino field 
$\tilde{\chi}^{\prime}_{\mbox{\tiny 6D}}=T\tilde{\chi}_{\mbox{\tiny 6D}}$.
Then the equations in terms of the new components of $\tilde{\chi}^{\prime}_{\mbox{\tiny 6D}}$ become first order differential equations of the radial coordinate $r$, 
so that one can easily solve the equations.

The generic form of these 24 equations can be written as follows: 
\begin{eqnarray}
{d \over d r}\tilde{\chi}_{\mbox{\tiny 6D}~m}^{\pm N}
-{m+n(N)\over r}\tilde{\chi}_{\mbox{\tiny 6D}~m}^{\pm N}
+{\alpha(N)\over h(r)^2}{d h(r) \over d r}\tilde{\chi}_{\mbox{\tiny 6D}~m}^{\pm N}=0, \label{equation-Y}
\end{eqnarray}
where $n(N)$ is an integer which depends on each $N$. 
The real numbers $\alpha(N)$ are found to be  
\begin{eqnarray}
+~:~~\alpha&=&
\{~2~,~{3\over2}~,~{3\over2}~,~{3\over2}~,~1~,~1~,~{7\over2}~,~-1~,~1~,~1~,~{7\over2}~,~{3\over2}~\} \n
-~:~~\alpha&=&
\{~1~,~{3\over2}~,~{3\over2}~,~{3\over2}~,~2~,~2~,~-{1\over2}~,~4~,~2~,~2~,~-{1\over2}~,~{3\over2}~\} 
\label{sets}
\end{eqnarray}
for each chirality.
These sets of numbers are exactly the same as the ones  
encountered  in the domain-wall type case \cite{Kimura:2009tb}. 
Note, however, that, unlike \cite{Kimura:2009tb}, our equations (\ref{equation-Y}) 
depends on the Fourier frequency $m$. 

The solutions of (\ref{equation-Y}), with a constant of integration $C$, are 
\begin{eqnarray}
\tilde{\chi}_{\mbox{\tiny 6D}~m}^{\pm N}=Cr^{m+n(N)}e^{\alpha\over h(r)}.
\end{eqnarray}
%
We consider the cases of $\alpha>0$ and $\alpha<0$ separately.

({\it i}) $\alpha(N)>0$\\
In this case the boundary condition is satisfied if and only if $C=0$,
and therefore there are no localized modes. 

({\it ii}) $\alpha(N)<0$\\
In this case, any Fourier mode satisfies the boundary condition. However, 
we also require that the mode must be normalizable $\int d^2x|\tilde{\chi}|^2<\infty$. 
Such modes are the ones with $m+n(N)=0$; only a single Fourier mode corresponds 
to a normalizable mode and is localized for each negative $\alpha(N)$. 

Therefore, the sets (\ref{sets}) show that there is only one normalizable localized 
mode with positive chirality, while there are two with negative chirality. This is 
exactly the same result as \cite{Kimura:2009tb}, confirming the claim that 
there exists one net chiral zeromode localized on this heterotic five-brane system.

%
%
%
\section*{Acknowledgments}
We thank Shun'ya Mizoguchi and Tetsuji Kimura for discussions and comments.


\end{document}